\documentclass[astrosymb]{aastex631}

\usepackage{tablefootnote}

\usepackage[T1]{fontenc}
\usepackage{amsmath}
\usepackage{epigraph}

\newcommand{\hide}[1]{}

\def\cm{\hbox{cm}}
\def\sec{\hbox{s}}
\def\f17{f_{17}}
\def\Mpc{\hbox{Mpc}}

\def\km{\hbox{km}}
\def\kms{\hbox{km s$^{-1}$}}

\def\Myr{\hbox{Myr}}
\def\Gyr{\hbox{Gyr}}

\def\kms{\hbox{km}\,\hbox{s}^{-1}}
\def\ergcm2s{\ifmmode {\rm\,erg\,cm^{-2}\,s^{-1}}\else
                ${\rm\,ergs\,cm^{-2}\,s^{-1}}$\fi}
\def\ergsec{\ifmmode {\rm\,erg\,s^{-1}}\else
                ${\rm\,ergs\,s^{-1}}$\fi}
\def\mstar{\ifmmode {M^*_{UV}}\else
                ${M^*_{UV}}$\fi}
\def\phistar{\ifmmode {\phi^*}\else
                ${\phi^*}$\fi}
\def\zo{\ifmmode {12 + \log[{\rm O}/{\rm H}]}\else  
                ${12 + \log[{\rm O}/{\rm H}]}$\fi}
                
\def\lya{\ifmmode {\hbox{Ly}\alpha}\else  
                Ly$\alpha$\fi}

\def\oIII{[\ion{O}{3}]4959,5007}
\def\oIIIa{[\ion{O}{3}]5007}
\def\oIIIb{[\ion{O}{3}]4959}
\def\oIIIc{[\ion{O}{3}]4363}
\def\oII{[\ion{O}{2}]3726,3729}
\def\oIIa{[\ion{O}{2}]3726}
\def\oIIb{[\ion{O}{2}]3729}
                
\def\JWSTobject{J072326-732656 (04590)}  
\def\JWSTobjtwo{J072322-732606 (06355)}  
\def\JWSTobjthree{J072320-732604 (10612)}  
\def\SMACS{SMACS~0723}
\def\SMACSfull{SMACS~0723.3-7327}

\shorttitle{Finding Peas in the Early Universe}
\shortauthors{Rhoads et al.}

\begin{document}

\title{Finding Peas in the Early Universe with {\it JWST}}

\author{James E. Rhoads}
\affil{Astrophysics Science Division, NASA Goddard Space Flight Center, 8800 Greenbelt Road, Greenbelt, Maryland, 20771, USA}
\email{James.E.Rhoads@nasa.gov}

\author{Isak G. B. Wold}
\affil{Astrophysics Science Division, NASA Goddard Space Flight Center, 8800 Greenbelt Road, Greenbelt, Maryland, 20771, USA}
\affil{Department of Physics, The Catholic University of America, Washington, DC 20064, USA }
\affil{Center for Research and Exploration in Space Science and Technology, NASA/GSFC, Greenbelt, MD 20771}

\author{Santosh Harish}
\affil{School of Earth and Space Exploration, Arizona State University, Tempe, AZ 85287, USA}

\author{Keunho J. Kim}
\affil{Department of Physics, University of Cincinnati, Cincinnati, OH 4521, USA}

\author{John Pharo}
\affil{School of Earth and Space Exploration, Arizona State University, Tempe, AZ 85287, USA}

\author{Sangeeta Malhotra}
\affil{Astrophysics Science Division, NASA Goddard Space Flight Center, 8800 Greenbelt Road, Greenbelt, Maryland, 20771, USA}
\email{Sangeeta.Malhotra@nasa.gov}

\author{Austen Gabrielpillai}
\affil{Astrophysics Science Division, NASA Goddard Space Flight Center, 8800 Greenbelt Road, Greenbelt, Maryland, 20771, USA}

\author{Tianxing Jiang}
\affil{School of Earth and Space Exploration, Arizona State University, Tempe, AZ 85287, USA}

\author{Huan Yang}
\affil{Las Campanas Observatory, Carnegie Institution of Washington, Casilla 601, La Serena, Chile}



\begin{abstract}
The Early Release Observations (EROs) of {\it JWST} beautifully demonstrate the promise of {\it JWST} in characterizing the universe at cosmic dawn.  We analyze the ERO spectra of three $z \sim 8$ galaxies to determine their metallicities,  gas temperatures and ionization. These galaxies offer the first opportunity to understand the physical properties of epoch-of-reionization galaxies through detailed rest-optical emission line spectroscopy.  We show that these objects have metal abundances $12+\log[O/H] \approx 6.9 - 8.2$, based on both the $T_e$ method and on a recent calibration of the $R_{23}$ metallicity indicator.  Since the spectra are some of the earliest science data from JWST, we compare several line ratios with values expected from robust physics, to validate our measurement procedures. We compare the abundances and emission line ratios to a nearby sample of Green Pea galaxies--- a population of nearby emission line galaxies whose UV properties resemble epoch-of-reionization galaxies, and which often have large Lyman continuum escape fractions. The JWST data show striking further similarities between these high redshift galaxies and nearby Green Peas.  The $z\sim 8$ galaxies span the metallicity range covered by Green Peas. They also show the compact morphology that is typical of emission line dominated galaxies at all redshifts.  Based on these similarities with Green Peas, it is likely that these are the first rest-optical spectra of galaxies that are actively driving cosmological reionization.
\end{abstract}

\epigraph{And as I fixed upon the down-turned face\\
That pointed scrutiny with which we challenge\\
The first-met stranger in the waning dusk\\
I caught the sudden look of some dead master\\
Whom I had known, forgotten, half recalled\\
Both one and many; in the brown baked features\\
The eyes of a familiar compound ghost\\
Both intimate and unidentifiable.}{\textit{T. S. Eliot\\Little Gidding}}




\section{Introduction} \label{sec:intro}
{\it JWST} was built to visit a time when galaxies were young.
In its first public release of scientific data, it has done exactly that, with rest-frame optical spectroscopy of three galaxies in the epoch of Cosmic Dawn (at redshifts $z>7$), as well as deep imaging of a cluster field, which provides extra magnification due to gravitational lensing. In one sense, these are automatically young galaxies, for they are observed when the universe itself was $\la 700$ Myr old, or about 5\% of its current age.

There are more specific markers of youth that derive directly from the properties of individual galaxies.   These include low stellar  mass; active star formation; and low abundances of the heavy elements produced by nuclear fusion in stars.  We will show that the most distant galaxy in the  {\it JWST} early release NIRSpec observations is young by all of these criteria.   The other two epoch-of-reionization galaxies in the same data set are still young objects by most criteria, albeit somewhat higher in both mass and metal abundance.


Detailed studies of galaxies at these redshifts will shed light on the reionization of intergalactic hydrogen, which was the landmark event of Cosmic Dawn--- the first time when bound objects had a global impact on the universe.  Combined constraints from the quasar spectroscopy \citep{Fan2006,Yang2020},  \lya\ statistics from both line-selected \citep{Malhotra2004} and continuum-selected \citep{Stark2010} galaxies, the cosmic microwave background \citep{Planck2016}, and ionizing photon production models \citep{Bouwens2015,Robertson2015,Finkelstein2019, Naidu2020} now suggest that the transition from neutral to ionized gas began before $z\approx 8.5$ and was largely finished by $z=6.5$.  Despite the expected obscuration of \lya\ by neutral gas \citep{miralda-escude1998}, \lya\ galaxies have been identified at redshifts up to $z\approx 8.6$ \citep{Zitrin2015}, and clusters or groups of \lya\ galaxies (indicating ionized bubbles) as far back as $z=7$ \citep{Castellano2016,Hu2021} to $7.7$ \citep{Tilvi2020}.   Wide area surveys for \lya\ emitters at $z\approx 7$ suggest that the IGM was in fact highly ionized by $z=7$ \citep{Itoh2018, Zheng2017, Hu2019, Goto2021, Wold2022}.  Observational constraints become much poorer at higher redshifts, due largely to the difficulty of infrared observations through Earth's atmosphere, and it is here that {\it JWST} is poised to drive rapid observational progress. 

To place these {\it JWST} Cosmic Dawn galaxies in context, we have measured their rest frame optical emission line ratios.   We have used the measurements to derive key physical parameters of the galaxies, notably including the gas phase oxygen abundance.   

Despite the extreme properties seen in these objects, they are not without close analogs in the nearby universe.  We compare the {\it JWST} line ratios to those of local emission line galaxy samples, including Green Pea galaxies at $z \la 0.3$ \citep{Cardamone2009, Jaskot2013, Henry2015, Izotov2016, Yang2016, Yang2017, Izotov2018, Jiang2019, Brunker2020, Kim2020, Kim2021} and extreme emission line objects selected at $z\sim 0.6$ \citep{Kakazu2007,Ly2016}.  We find that the most extreme of the {\it JWST} sources has a metal abundance $\zo \sim 7.0$.  Within present uncertainties, it may or may not be the most metal poor galaxy known.   It is clear that galaxies of comparably low metallicity {\it do} exist in the nearby universe.  But, it is equally clear that  such objects are exceedingly rare at low redshift.  They account for $\la 1\%$ of Green Pea galaxies 
despite preselection  using criteria, e.g. the presence of strong nebular emission lines, that should enhance their representation in that sample.

Regardless of what future data sets tell, this is a huge step forward.  The first release of scientific data from {\it JWST} has enabled us to apply powerful tools that could previously be used only with great difficulty beyond $z\sim 1$, and even then only up to $z\sim 3$.

In this paper, we discuss the {\it JWST} observations (sec.~\ref{sec:obs}), line ratio measurements (sec.~\ref{sec:measurements}), and consistency checks applied to the measurements (sec.~\ref{sec:checks}); and also the selection of comparison objects from {\it Sloan Digital Sky Survey} Green Pea galaxies (sec.~\ref{sec:gp}).  We then present our analysis of the physical conditions drawn from the line ratio measurements (sec.~\ref{sec:analysis}).  Finally, we discuss the implications of the results (sec.~\ref{sec:discuss}), and summarize our conclusions (sec.~\ref{sec:conclusions}). 
Throughout the paper, where relevant, we use a flat concordance cosmology with $H_0 = 69.6 \km\,\sec^{-1}\,\Mpc^{-1}$, $\Omega_{tot} = 1$, $\Omega_m = 0.286$, and $\Omega_\Lambda = 0.714$.  

There is a rapidly growing body of work on the {\it JWST} Early Release Observations, including papers that examine the properties of the same three galaxies (\JWSTobject, \JWSTobjtwo, and \JWSTobjthree) that are our focus here.  These include \citet{Carnall2022}, which presents redshifts, stellar masses, and star formation histories of these galaxies; \citet{Adams2022}, which recovers these three sources among a Lyman break search for $z\ga 9$ objects; and \citet{Schaerer2022}, which presents a metal abundance analysis with notable similarities to the present work.   While most of our analysis was carried out before those works became available, we will compare our conclusions to theirs in section~\ref{sec:discuss}.

\section{Observations and Data Analysis} \label{sec:obs}

\subsection{JWST observations}
We used the {\it JWST} Early Release Observations of the galaxy cluster \SMACSfull\ (hereafter \SMACS), and especially the NIRSpec multi-object spectroscopy (program JW02736; observations 007 and 008).  This consisted of a single configuration of the NIRSpec multi-shutter array, observed with a range of blocking filters and gratings.   The most critical instrumental setup for this paper was with the F290LP blocking filter and G395M grism, which covers $\lambda_b \approx 2.9 $ to $\lambda_r \approx 5.2 \mu$m with a resolving power $R\sim 1000$.   The NRSIRS2 detector readout pattern was used with 20 groups, 2 integrations, and a 3 shutter slitlet nod pattern.  For our targets of interest (\JWSTobject, \JWSTobjtwo, \JWSTobjthree) this configuration resulted in on-target exposures of $8.8$ ks for both observations (o007 and o008).

\subsection{Emission line ratio measurements} \label{sec:measurements}
We downloaded the Level 3 spectra for the SMACS J0723.3-7327 program from the STScI MAST server.  These are pipeline calibrated products, using a mixture of ground and on-orbit calibration data \citep{Rigby2022}.  We base our analysis on the 1d spectra for each of two long exposures (o007 and o008).   We measured the fluxes for several detected emission lines in each spectrum, notably including \oIII, \oIIIc, and \oII, and the Balmer series lines of hydrogen from H$\beta$ through $H\epsilon$.  

\hide{
For each line, we integrated the flux by a direct sum over a window of width $\delta \lambda / \lambda = \pm 0.001$ (or equivalently $\pm 300 \kms$) around the line central wavelength, typically corresponding to four pixels in the 1D spectra.  We accounted for the mixed unit system in the L3 products (correcting the $\lambda f_\nu$ units that a straight integration would yield into energy flux units before forming line ratios).
We estimated the continuum level for each line using the median pixel value in a window with $0.001 <  |\delta \lambda / \lambda | < 0.006$ after rejecting outliers, and subtracted this continuum from each line flux.}

To measure the line fluxes, we fitted Gaussian profiles to the 1D spectrum.  To obtain robust fitting results on both strong and weaker lines, we performed simultaneous fits of nearby line sets, so that the wavelength and line width were effectively fixed by well detected lines, allowing more confident extraction of the weak lines.   Finally, we formed line ratios separately for each of the two exposures, and combined the two line ratio estimates using an inverse variance weighting.  The \oIIIb\ line of \JWSTobject\ in the L3 exposure o008 product downloaded around July 15, 2022 appears to have some corrupted pixels, which result in a strongly negative flux measurement.   We chose to replace the \oIIIb\ line flux in this instance with the value expected from the theoretical \oIIIb/\oIIIa\ ratio, given that the \oIIIa\ line is well measured and the  \oIIIb/\oIIIa\ observed in exposure o007 is consistent with the theoretical value.  

To estimate errors on line flux ratios, we first noted that the RMS of the continuum flux in the Level~3 1D spectra exceeded the noise reported in the L3 spectra, by a factor of $\sim 1.5$ to 2.   We therefore applied a $2\times$ correction to the noise levels input to the line fitting software (MPFITFUN in IDL), and thereafter used the reported uncertainties in fitted line parameters to determine random line flux errors.  Finally, we applied a noise floor of 5\% to measured line fluxes (added in quadrature with the random flux errors reported by MPFITFUN), to allow for low level wavelength dependent calibration errors or similar systematic problems.

The line flux ratios measured through this procedure are summarized in table~\ref{tab:jw_ratios}, along with derived
quantities including electron temperature and gas-phase oxygen abundances (see section~\ref{sec:analysis} below).

\begin{deluxetable}{ccccccccccc}




\tablecaption{Observed and derived properties of the {\it JWST} spectroscopic sample and of two local analogs.}

\tablenum{1}

\tablehead{\colhead{ID} & \colhead{Spec. ID} & \colhead{Redshift} & \colhead{4363/OIII} & \colhead{OIII/Hb} & \colhead{OII/Hb} & \colhead{R23} & \colhead{O32} & \colhead{Z(R23)\tablenotemark{a}} & \colhead{Te(OIII)} & \colhead{Z(Te)\tablenotemark{a}} \\
\colhead{} & \colhead{} & \colhead{} & \colhead{} & \colhead{} & \colhead{} & \colhead{} & \colhead{} & \colhead{} & \colhead{$10^4$ K} & \colhead{} 
} 
\startdata
J072326-732656 & 04590 & 8.495 & 0.058$\pm$0.014 & 3.92$\pm$0.33 & 0.30$\pm$0.11 & 4.18$\pm$0.38 & 12.26$\pm$4.25 & 7.2 & 3.72$\pm$0.99 & 6.88$\pm$0.15   \\
J072322-732606 & 06355 & 7.664 & 0.011$\pm$0.003 & 9.46$\pm$0.63 & 1.19$\pm$0.12 & 10.72$\pm$0.70 & 7.89$\pm$0.61 & 8.2 & 1.34$\pm$0.16 & 8.09$\pm$0.16   \\
J072320-732604 & 10612 & 7.659 & 0.029$\pm$0.011 & 11.92$\pm$2.06 & 0.71$\pm$0.28 & 12.67$\pm$2.23 & 17.29$\pm$4.49 & 8.2 & 2.19$\pm$0.54 & 7.68$\pm$0.24   \\
\hline
J082701+342951 & SDSS & 0.0854 &  0.033$\pm$0.003 &  8.03$\pm$0.60 &  0.38$\pm$0.04 &  8.41$\pm$0.61 & 20.85$\pm$1.64 & 7.7 & 2.40$\pm$0.14 & 7.44$\pm$0.05 \\
J122051+491555 & SDSS & 0.0123 &  0.028$\pm$0.003 &  4.02$\pm$0.28 &  0.25$\pm$0.05 &  4.27$\pm$0.30 & 16.23$\pm$3.40 & 7.25 & 2.12$\pm$0.15 & 7.24$\pm$0.06 \\
\enddata

\tablenotetext{a}{Gas phase oxygen abundances tabulated above are given in units of $12 + \log(O/H)$.  Solar oxygen abundance in these units is approximately 8.7.}


\label{tab:jw_ratios}
\end{deluxetable}

\begin{figure}
\includegraphics[width=\linewidth]{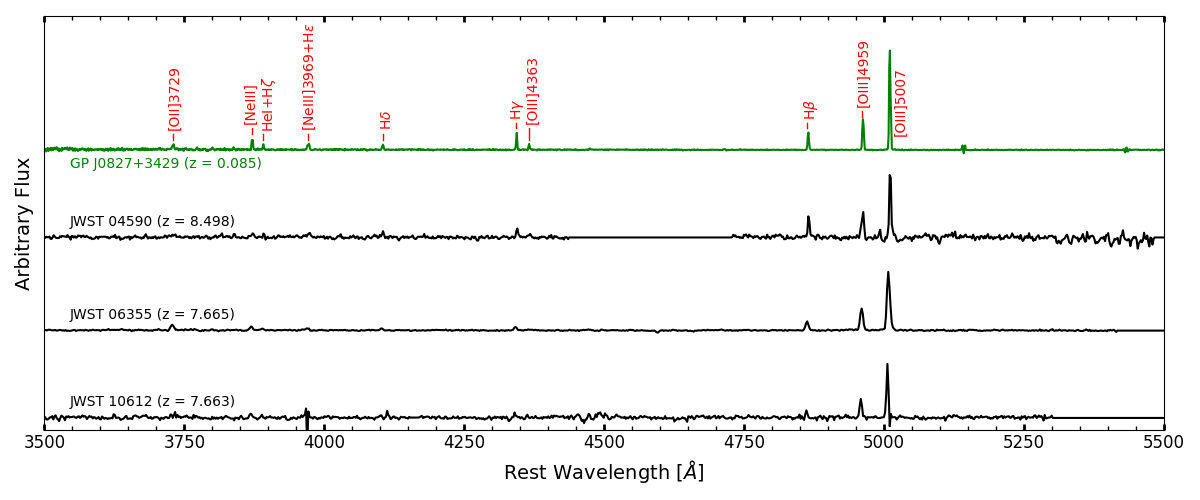}
\caption{Comparison of the rest-optical spectra of the $z\approx 0.085$ Green Pea galaxy J0827+3429 and the epoch-of-reionization galaxies \JWSTobject, \JWSTobjtwo, and \JWSTobjthree.   
The Green Pea galaxy shown is selected as one of two best analogs to the line ratios of \JWSTobject\ among the Green Pea sample from \citet{Yang2019} and \citet{Jiang2019} (see section~\ref{sec:gp}).
For display purposes we have masked a few spurious features in these early-release JWST spectra that did not match any known spectral line, after confirming that these features were not consistently present in both exposures.
\label{fig:gp_jwst_compare}}
\end{figure}

\subsection{Consistency Checks and Robustness} \label{sec:checks}
Given that we are doing science with pipeline-processed early-release data, it is likely that there will be future refinements to the data processing and possible that these may have impact on relevant measurements.  We have taken several steps to test for robustness and mitigate the impact of possible issues in data processing.   

First, as discussed above, we renormalized the stated flux errors in the L3 spectra to bring them into line with the observed RMS continuum flux in the 1D spectra.

Second, we treated each exposure separately in measuring fluxes and forming flux ratios, and combined results statistically thereafter.   This means that any overall flux normalization issue between the two exposure on each object will not impact the results.  Moreover, we used only line fluxes, after  subtracting a locally determined continuum level, so that our results are insensitive to any slowly varying additive component in the spectra.


We checked our measured ratios against theoretical expectations for  those line ratios with values largely determined by atomic physics, namely
\oIIIb/\oIIIa, where theory predicts  0.335 \citep{Dimitrijevic2007};  H$\gamma$/H$\beta$, where theory predicts 0.474 \citep{Osterbrock2006}; and H$\delta$/H$\beta$, where theory predicts 0.262  \citep{Osterbrock2006}.
The measured ratios are generally within 1--2$\sigma$ of their theoretical values.  Results of these checks are given in table~\ref{tab:jw_checks}.

\begin{table}
\begin{tabular}{llllllll}
Object & exposure & 4959/5007 		& Deviation	&  Hgamma/Hbeta     & Deviation &   Hdelta/Hbeta & Deviation  \\
\hline
Theory &     -    &    0.3356  & 0		&  0.474   & 0        & 	0.262  & 0 \\
\hline
04590 &	  o007	  &   0.306 $\pm$ 0.050 & -0.6$\sigma$ &  0.679 $\pm$ 0.149 & 1.4$\sigma$ & 0.431 $\pm$ 0.155 & 1.1$\sigma$  \\
04590 &	  o008	  &   - corrupted - 	& - & 0.552 $\pm$ 0.101 & 0.7$\sigma$     & 0.348 $\pm$ 0.338 & 0.2$\sigma$  \\
04590 &	  combined &  0.306 $\pm$ 0.050	& -0.6$\sigma$ & 0.592 $\pm$ 0.084 & 1.4$\sigma$  & 0.417 $\pm$ 0.141 & 1.1$\sigma$  \\
\hline
06355 &	  o007 	  &   0.382 $\pm$ 0.030 &  1.5$\sigma$ &    0.523 $\pm$ 0.082 & 0.6$\sigma$  &  0.310 $\pm$ 0.318 & 0.2$\sigma$\\
06355 &	  o008    &   0.371 $\pm$ 0.031 &  1.2$\sigma$ &    0.464 $\pm$ 0.077 & -0.2$\sigma$ & 0.258 $\pm$ 0.256  & 0.0$\sigma$  \\
06355 &	  combined &  0.377 $\pm$ 0.022 &  1.9$\sigma$ &    0.490 $\pm$ 0.056  & 0.3$\sigma$ &  0.278 $\pm$ 0.198 & 0.1$\sigma$ \\
\hline
10612 &	  o007     &  0.313 $\pm$ 0.039 &  -0.6$\sigma$ &   0.842 $\pm$ 0.391 & 0.9$\sigma$ &   0.194 $\pm$ 0.351 & -0.2$\sigma$   \\
10612 &	  o008	  &   0.385 $\pm$ 0.053 &  0.9$\sigma$  &   0.726 $\pm$ 0.361 & 0.7$\sigma$ &  2.04 $\pm$ 2.40  & 0.8$\sigma$ \\
10612 &	  combined &  0.338 $\pm$ 0.031 &  0.04$\sigma$ &   0.779 $\pm$ 0.264 & 1.1$\sigma$ &  0.232 $\pm$ 0.230 & -0.1$\sigma$ \\
\end{tabular}
\caption{Table of line flux ratio consistency checks.  ``Deviation'' indicates the discrepancy between the observed and theoretical ratio, in $\sigma$ units.
\label{tab:jw_checks}}
\end{table}


Finally, we note that our key physical conclusions (section~\ref{sec:analysis}) depend primarily on pairs of lines that are relatively close together in wavelength. Because \oII\ is comparatively weak in all of the objects studied here, its impact on the inferred metallicity is modest. Therefore, the widest wavelength range that strongly affects results is the span from \oIIIc\ to \oIIIa, a factor of $\sim 1.15$ in wavelength.  This range is covered by a single NIRSpec configuration (although the two lines fall on different detectors in object \JWSTobject).  The consistency checks above include the ratios H$\gamma/$H$\beta$ and H$\delta/$H$\beta$, which span a similar wavelength interval.

\subsection{SDSS Green Pea comparison} 
\label{sec:gp}
\begin{figure}
\includegraphics[width=0.495\textwidth]{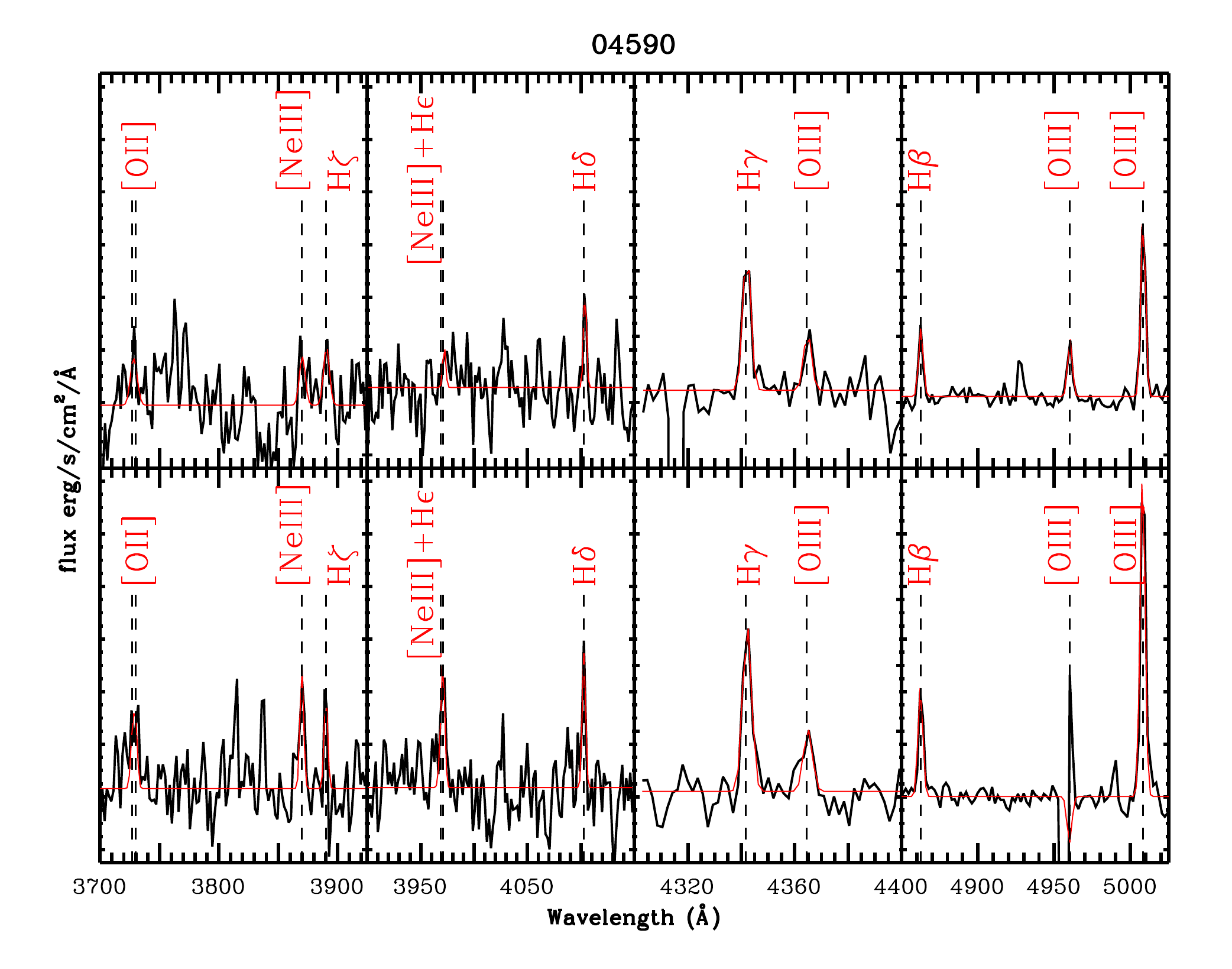}
\includegraphics[width=0.495\textwidth]{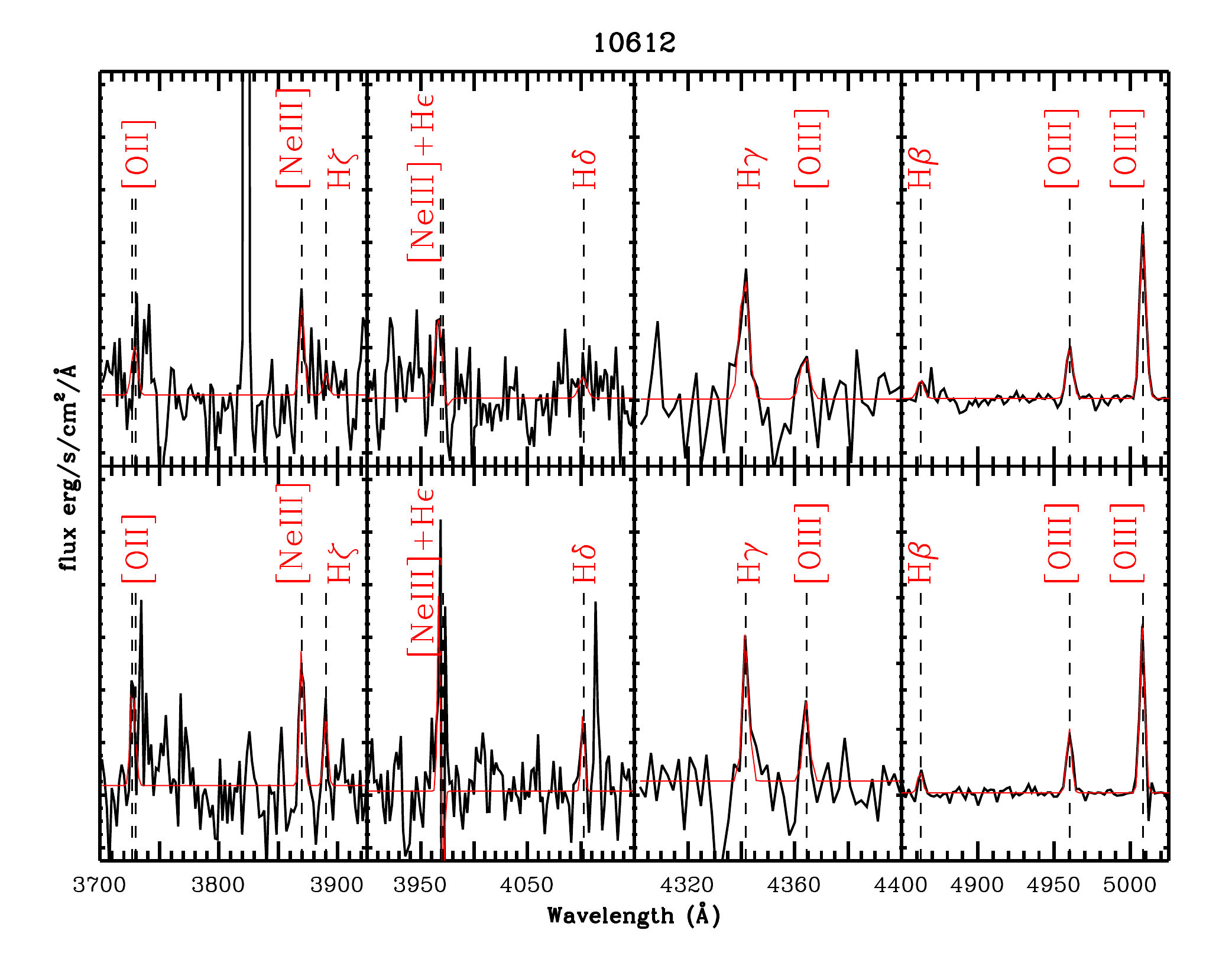}\\
\includegraphics[width=0.495\textwidth]{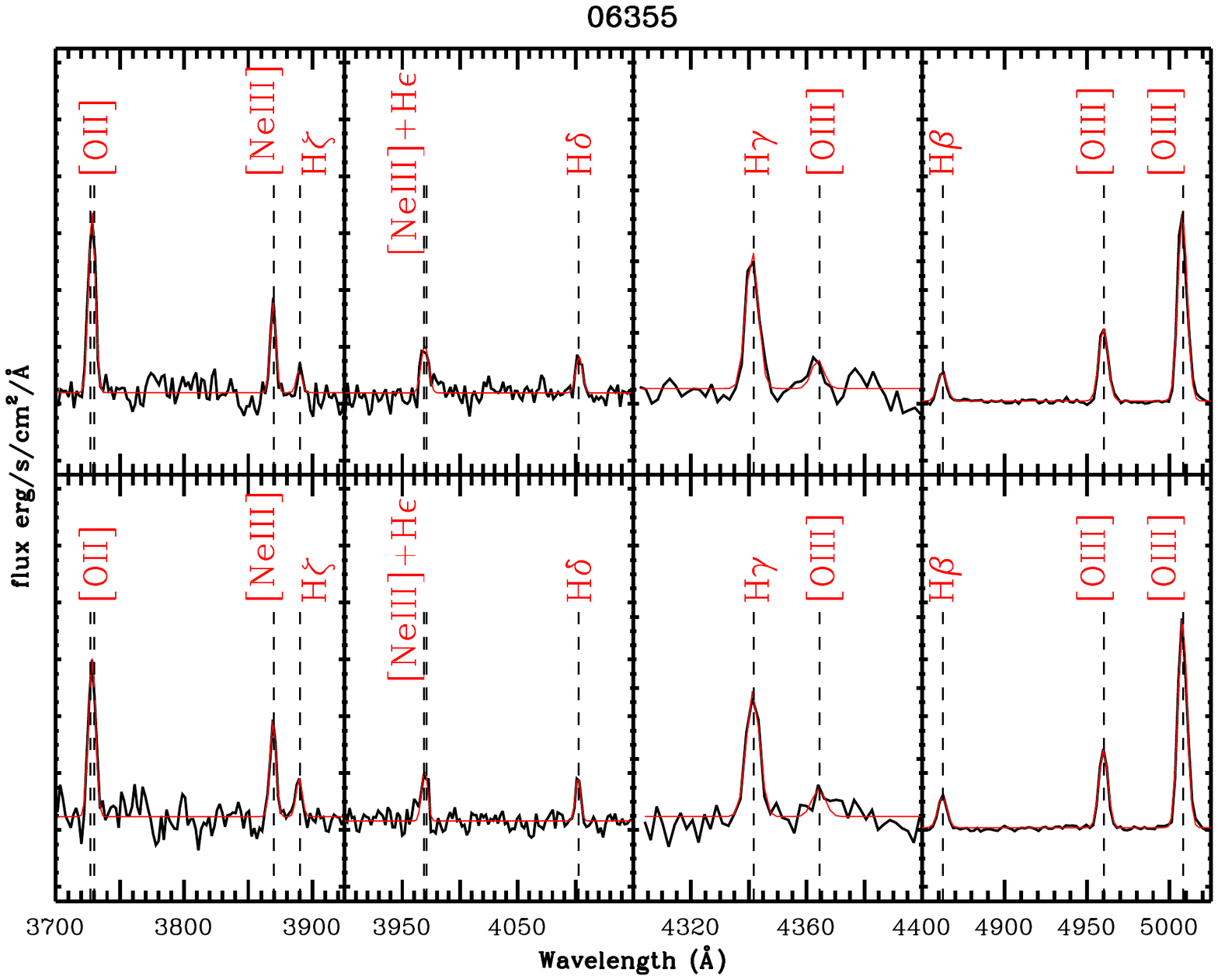}
\includegraphics[width=0.495\textwidth]{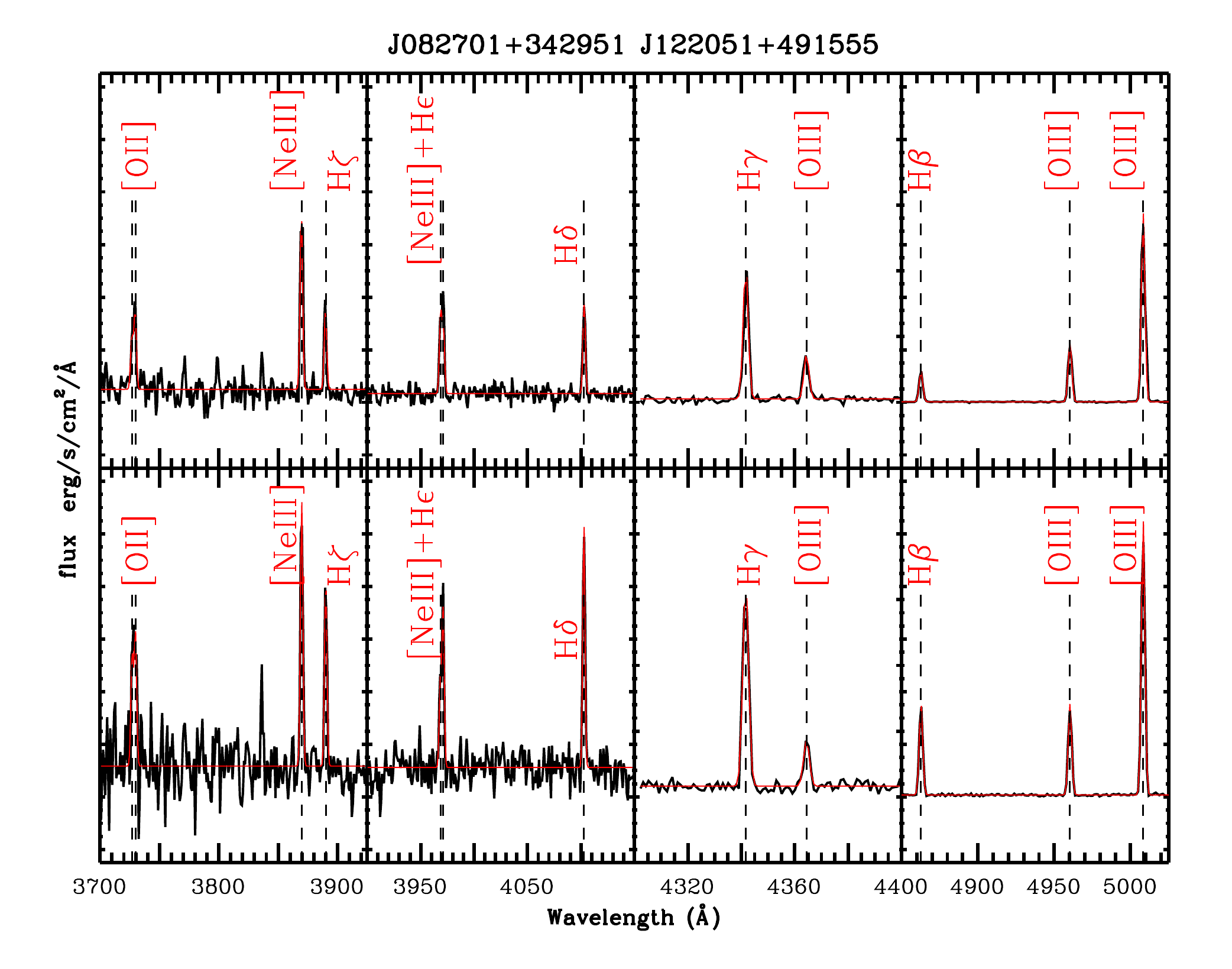}
\caption{The emission lines and fits to these lines for objects in our analysis are shown here.  The four panels correspond to \JWSTobject (top left), \JWSTobjthree\ (top right), \JWSTobjtwo\ (bottom left), and two SDSS Green Pea spectra (bottom right).  Within each of the JWST panels, the upper and lower rows of sub-panels correspond to the two distinct NIRSpec exposures.  For the Green Pea panel, upper and lower sub-panels correspond to two distinct objects.   Within each row, four distinct wavelength segments are plotted.  Each segment shows a linked line set, where we performed a joint fit to 2--4 emission lines.  Each fit included three parameters to describe the redshift, line width, and continuum level, plus one parameter per emission line to measure line amplitudes.
The four fitted line sets are (from blue to red)
[OII] 3727, 3729 + [NeIII] 3869 + H$\zeta$ 3890; 
[NeIII] 3968 + H$\epsilon$ 3971 + H$\delta$ 4102; 
H$\gamma$ 4341 + [OIII] 4364; 
and H$\beta$ 4862 + [OIII] 4960 + [OIII] 5008.
\label{fig:linefits}}
\end{figure}

The spectrum of \JWSTobject\ in particular is reminiscent of some of the most extreme emission line galaxies in the nearby universe.   To explore this, we searched the database of Green Pea galaxies from \citet{Yang2019,Jiang2019} to find galaxies with similar line ratios.
\hide{SELECT ...   FROM green_peas AS gp
where oiii4363_observed/oiii4363_err > 2 AND oiii4363_observed/oii3727_observed > 0.5 AND ((1.3*oiii5007_observed+oii3727_observed)/hbeta_observed)<8 AND (1.3*oiii5007_observed/oii3727_observed)>10 AND oiii4363_observed/oiii5007_observed> 0.035} 
Specifically, we required $S/N($\oIIIc$) > 2$, $f($\oIIIc$) / f($\oIII$) > 0.035$, $f($\oIIIc$) / f($\oII$) > 0.5$,  $R_{23} < 8$, and $O_{32} > 10$.   This yielded two objects,  J082701+342951
and J122051+491555, from a parent database of about 1000 sources (which had been previously selected on a minimum equivalent width threshold in \oIII\ and/or H$\beta$).   We compare the rest-frame spectra of GP J082701+342951 and the three JWST sources in figure~\ref{fig:gp_jwst_compare}, and show both these green peas in the lower right panel of figure~\ref{fig:linefits}.   Spectral data are taken from SDSS Data Release DR12 \citep{Alam2015}.

The similarities in most line ratios are quite striking, especially between \JWSTobject\ and GP~J122051+491555, where all ratios apart from \oIIIc/\oIII\ are within $1\sigma$ (see table~\ref{tab:jw_ratios}).  This similarity gives new confidence to the conclusion that Green Peas are good analogs for epoch of reionization galaxies.

\section{Analysis} \label{sec:analysis} 

We measured gas-phase oxygen abundances using multiple approaches as described in the following sections.

\subsection{R23 method}
\begin{figure}
{\includegraphics[width=0.8\textwidth]{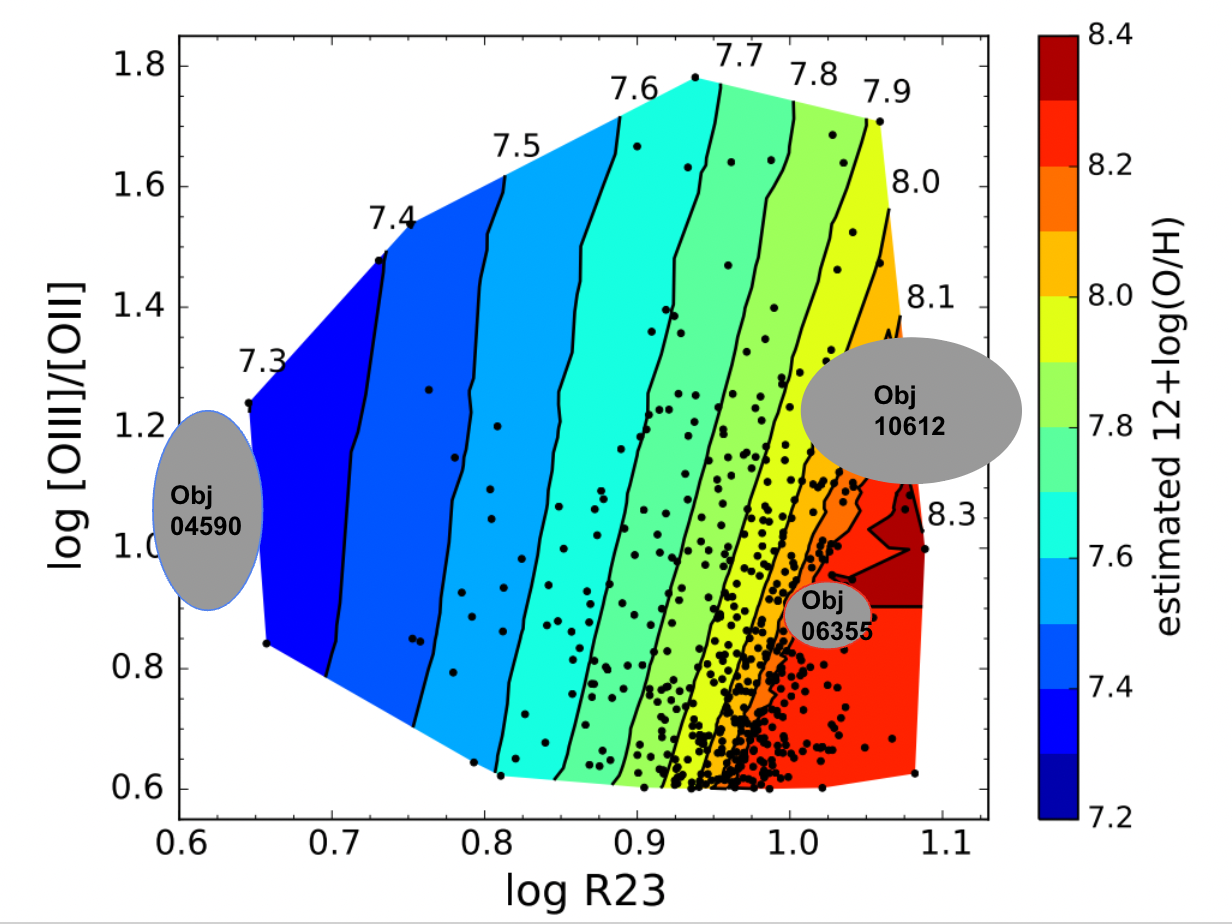}
\caption{This figure, based on figure 8 of \citet{Jiang2019}, shows the relation between $R_{23}$, $O_{32}$, and gas phase metallicity. Grey ovals mark the three $z>7$ objects, which have inferred $R_{23}$ metallicities of $\sim 7.2$ (\JWSTobject), $\sim 8.2$ (\JWSTobjtwo), and $\sim 8.2$ (\JWSTobjthree).   The first two measurements are in good agreement with the $T_e$ method results, while \JWSTobjthree\ appears to be an outlier in this relation.
Together, this trio of sources spans the range of $R_{23}$ seen in the comparison sample of Green Pea galaxies.}
\label{fig:txj}}
\end{figure}

First,  we apply the $R_{23}$ metallicity method using the calibration by
\citet{Jiang2019}.   This calibration is based on a sample of green pea galaxies, whose spectra have notable similarities to the spectrum of \JWSTobject (see figure~\ref{fig:gp_jwst_compare}). 
Direct interpolation off the $Z(R23, O32)$ relation yields a metallicity estimate $\zo \approx 7.2$ for \JWSTobject, and $8.2$ for \JWSTobjtwo\ and \JWSTobjthree\ (see figure~\ref{fig:txj}).

\subsection{Direct method}
\begin{figure}
{\includegraphics[width=0.8\textwidth]{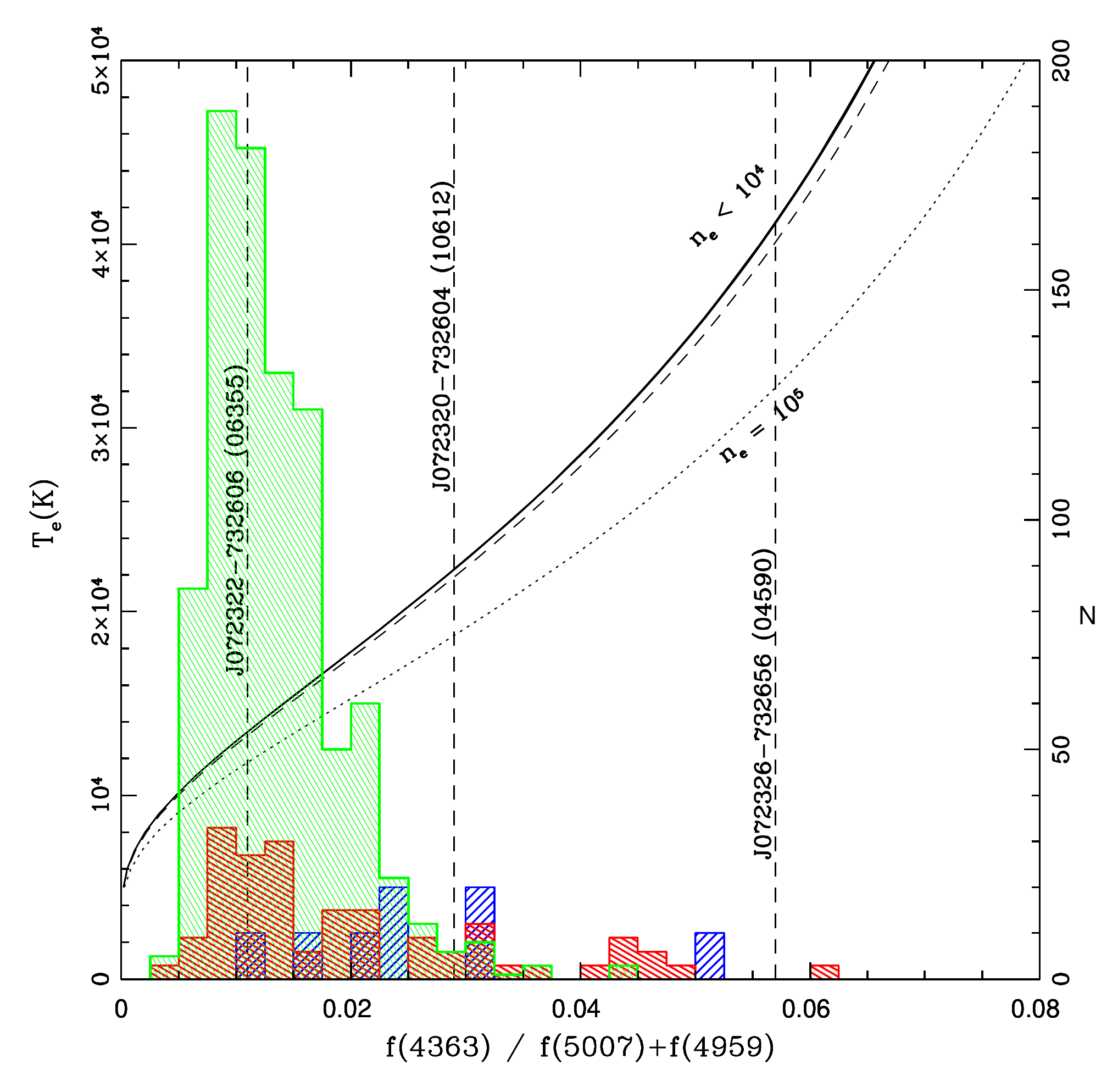}}
\caption{The electron temperature of [OIII] line emitting gas
is plotted as a function of the f(OIII 4363)/f(OIII 4959,5007) ratio, following the treatment
in \citet{Osterbrock2006}.  The solid curve shows the relation in the low-density limit, approximately $n_e < 10^3 \cm^{-3}$.  
The dashed curve shows the relation for $n_e = 10^4  \cm^{-3}$,
and the dotted curve the relation for $n_e = 10^5  \cm^{-3}$.
A vertical dashed line shows the measured value of this
ratio for  \JWSTobject.  Based on this, we infer an electron
temperature near 30,000K in the [OIII] region of this galaxy.
Histograms of comparison samples show that the f(OIII 4363)/f(OIII 4959,5007) ratio and the electron temperature in \JWSTobject\ 
are extreme but not unprecedented in lower-redshift samples.  The green histogram marks the $z\la 0.3$ Green Pea galaxy sample of \citet{Yang2019,Jiang2019}.   The blue and red histograms show intermediate redshift emission line selected samples from \citet{Kakazu2007} and \citet{Ly2016} (with numbers multiplied by 10 and 3, respectively, to display on the same y-axis as the GP sample).  All comparison samples have been restricted to objects with $S/N \ge 2.5$ in the \oIIIc\ line.
\label{fig:osterbrock}}
\end{figure}

Second, we use the well detected [OIII]4363\AA\ line to apply the $T_e$ (``direct'') method. 
Here we follow the method described in \citet{Jiang2019}, which follows prior work by \citet{Izotov2006}.    The method uses the ratio f(\oIIIc)/f(\oIII) to measure the electron temperature in the \oIII\ emitting gas.  Combining this temperature with the line flux ratio f(\oIII)/f(H$\beta$) furnishes an estimate of the O$^{++}$/H ratio.   Similar methodology yields the O$^+$/H ratio.  
Figure~\ref{fig:osterbrock} shows the relation between the observed f(\oIIIc)/f(\oIII)  ratio and temperature, which is inverted numerically.
The resulting value of $T_e(OIII) \approx (3.7 \pm 1) \times 10^4$K for \JWSTobject is extreme, though not without precedent among lower-redshift samples in the literature  \citep[e.g.,][]{Kakazu2007}.   
The mass ratios of $O^+/H$ and $O^{++}/H$ are determined using fitting formulas presented by \citet{Izotov2006}.  The temperature of the \oII\ emitting gas is not directly constrained, and 
past work has often used fitting formulas to estimate $t_2 \equiv T({\rm OII})/10^4 K$ based on $t_3 \equiv T({\rm OIII}) / 10^4 $ (e.g., \citet{Jiang2019} uses the $T({\rm OII})$ - $T({\rm OIII})$ relation from \citet{Izotov2006}, which is in turn based on a suite of photoionization models from \citet{Stasinska1990}).  We have used a modification of the fitting formula from \citet{Izotov2006}:
$t_2 =  -0.577 + t_3 \times (2.065 - 0.498 t_3)   \hbox{ for  } t_3  < 2.07$, and $t_2 = 1.562    \hbox{ for  } t_3  > 2.07 $.  This avoids an unphysical decrease of $t_2$ with increasing $t_3$.  Results (both $T_e$ and \zo) for all three {\it JWST} targets and the two Green Pea comparison objects are reported in table~\ref{tab:jw_ratios}, and span a range  $6.9 \la \zo \la 8.1$ that is similar to the range seen in $\sim 1000$ Green Peas \citep{Jiang2019}.

\subsection{CLOUDY modeling for \JWSTobject} 
\label{sec:cloudy}
To explore the physical conditions in the most extreme of the JWST sources, \JWSTobject, we have used the CLOUDY photoionization code\footnote{CLOUDY v. 17.03} \citep{Ferland2017} and the pyCLOUDY package \citep{Morisset2013} to calculate expected emission line ratios for a young, low-metallicity starburst. 
This package takes an input incident radiation spectrum and computes the full radiative transfer through a surrounding gas cloud, thereby predicting the resultant nebular emission spectrum.  
We adapt the model prescriptions used in \citet{Byler2020}, an analysis of UV and optical emission line diagnostics of metallicity and SF in young star-forming galaxies. Given the extreme observed line ratios, we adopt the minimum inner radius ($R_{inner} \sim 0.1$ pc) and maximum ionizing photon production ($Q_H \sim 10^{51}$ s$^{-1}$, comparable to a young star cluster) from the Byler grids, such that the model will have a high ionization parameter, $U$. $U$ is also dependent on the hydrogen density $n_H$, and the photoionization also depends on the shape of the incident spectrum. To quickly approximate the spectrum of a young burst of low-metallicity stars, we use a blackbody spectrum with $T_{eff}=60000$ K \citep[comparable to CLOUDY models used in high-excitation starbursts used in e.g.,][]{Steidel2014,Sanders2016}. 

\begin{figure}
{\includegraphics[width=0.8\textwidth]{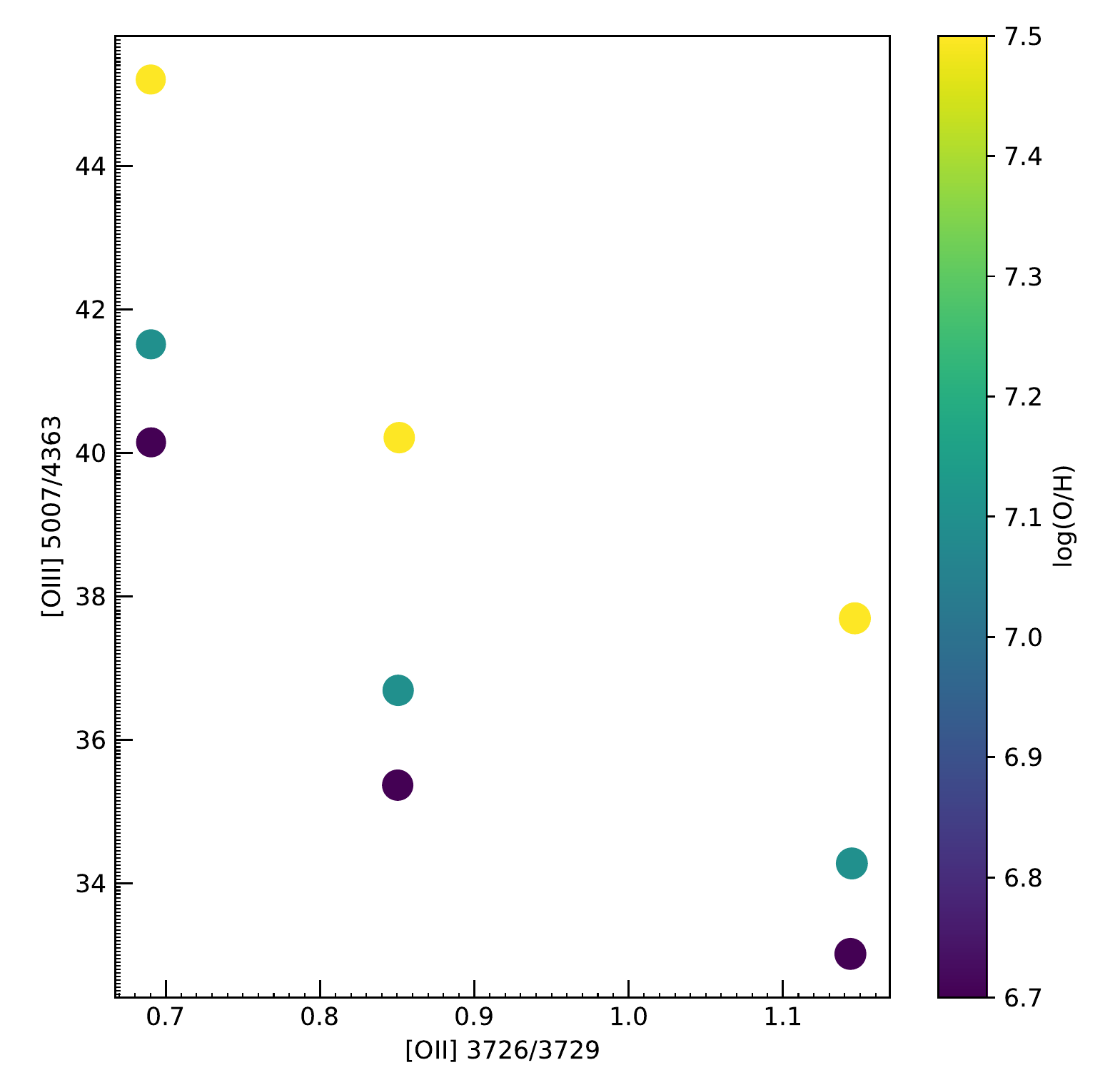}}
\caption{CLOUDY model results, showing the line ratios of \oIII\ to \oIIIc\ and \oIIa\ to \oIIb\ for a grid of models with electron densities of  $10^{1.5}$, $10^{2.5}$, and $10^3 \cm^{-3}$ [from left to right], and gas phase oxygen abundances $\zo = 7.5$, $7.1$, $6.7$ [from top to bottom].  Despite examining intense and low-metallicity conditions, none of the models tested reproduce the \oIIIc/\oIII\ ratio observed in \JWSTobject\ (although they match the value for \JWSTobjthree\ and easily exceed that for \JWSTobjtwo, which could be matched with less extreme models).  More detail on the models is presented in section~\ref{sec:cloudy}, and discussion of the implications is in section~\ref{sec:discuss}.
\label{fig:CLOUDY}}
\end{figure}

We then run CLOUDY with these settings with grids of metallicity ($6.7 < \log(12 + O/H) < 7.5$) and electron density ($1.5 < \log(n_e) < 3$), with ranges based on estimates from the observed emission line ratios as described above (probably). As the line ratios suggest low oxygen abundances in the nebular gas, in addition to scaling the oxygen abundance, we reduce other heavy elements in the model as well, scaling their values with the oxygen abundance. 
The predicted oxygen line ratios from this model grid are shown in Figure \ref{fig:CLOUDY}. Each ``column'' of model points represents a given density, with the left column most closely matching the density predicted from the observed OII ratio ($\sim 0.7$, corresponding to lower densities). The color of the points gives the gas-phase metallicity.

For the model to get close to the observed [O\textsc{iii}]5007/[O\textsc{iii}]4363 ratio, some combination of extremely low metallicity, high electron density, harder incident ionizing spectrum (as from e.g. particularly low stellar metallicity in the starburst), and compact cloud morphology is necessary.
However, {\it none} of the models tested actually reproduces the f(\oIIIc)/f(\oIII) ratio observed in \JWSTobject.  We consider possible explanations in section~\ref{sec:discuss} below.

\subsection{Size and Surface Brightness Comparisons}
\begin{figure}
\includegraphics[width=0.99\textwidth]{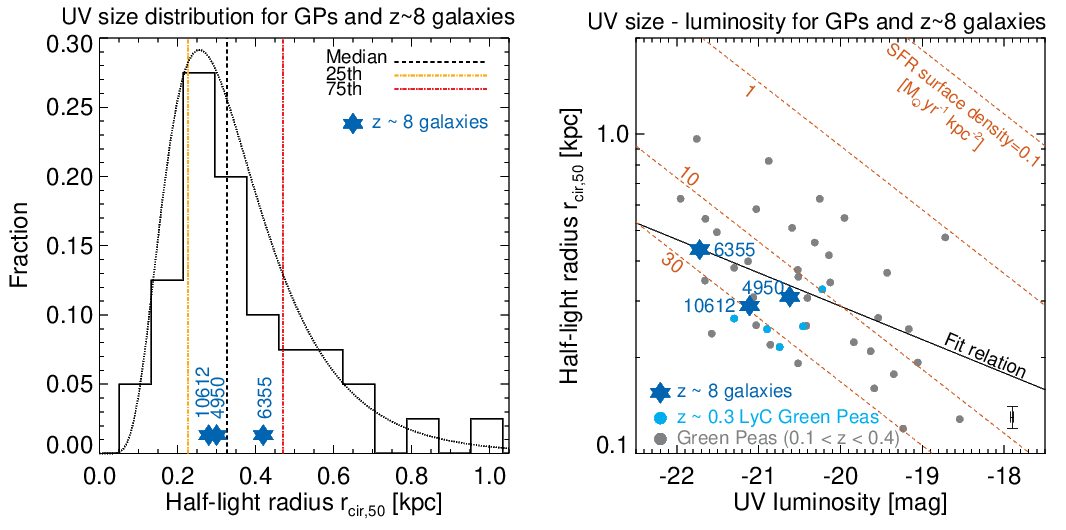}
\caption{Blue star-shaped symbols mark the sizes and magnitudes  of the {\it JWST} $z\sim 8$ galaxies compared to local analogs.   Left:  Histogram of Green Pea sizes (based on \citet{Kim2021}).   All three {\it JWST} sources lie in the 2nd or 3rd quartile of Green Pea sizes.   
Right: Absolute magnitude vs. size, with contours of constant star formation intensity \textbf{(i.e., orange dashed lines)} marked.   \textbf{Gray symbols} mark Green Peas.  Pale blue symbols mark known Lyman continuum leakers \citep{Izotov2016}.  The {\it JWST} sources populate the region of parameter space with the highest star formation intensity, which is also the region where known Lyman continuum leakers occur.
\label{fig:size}}
\end{figure}

Another characteristic of the local and low redshift emission line galaxies is their compact size and high star-formation rate per unit area (a.k.a, star formation surface density, $\Sigma$SFR). This could be due to the dominance of one compact star-forming region (or cluster) in the galaxy. We thus examined the sizes and star-formation intensities (Star-Formation Rate per unit area $\Sigma$SFR) for these galaxies. In figure~\ref{fig:size}, we show how the sizes of the three high redshift sources compare with low redshift Green Peas \citep{Kim2021}. The sizes plotted are half-light radii as derived in Source Extractor \citep{Bertin1996}. The directly measured circularized half-light radii (i.e., corrected for the galaxy ellipticity) range from {0.28 to 0.42 kpc}. If we apply a rough correction for lensing amplification due to the foreground galaxy cluster, using the reported magnification values from \citet{Carnall2022} which were obtained from the lens model provided by the RELICS team \citep{Coe2019}, the sizes drop to 0.1--0.26 kpc.
These lensing corrections are modest for two of the sources, with the exception being an  $\sim 10$ and a corresponding reduction in circularized radius of $\sim 1/3 \times$ for \JWSTobject (04590).
These sizes are typical of Green Peas ($\sim$ 0.33 kpc) at low redshifts, where we have measured them from well resolved near-UV images \citep{Kim2021}. 

We also calculate average star-formation intensities (SFI, equivalent to $\Sigma \rm{SFR} \equiv \frac{\rm{SFR}}{2 \pi r_{\rm{cir,50}}^{2}} \left (\frac{M_{\sun} \ \rm{yr}^{-1}}{{\rm kpc}^2} \right )$) for these three sources. The star-formation rate is calculated using observed rest-frame UV-continuum in F150W.   We apply no dust correction.  If dust corrections are in fact required, we could be under-estimating the SFI in these sources. We note that the Balmer line ratios in table~\ref{tab:jw_checks} do not require reddening, though this is not an especially strong constraint without an H$\alpha$ measurement.  
To convert the measured UV luminosity to the corresponding SFR in a consistent manner with the compared Green Peas, we apply the same conversion method as in \citep{Kim2021}. 
we adopt the solar bolometric magnitude of 4.74 \citep{Bessell1998}, and the UV to bolometric luminosity ($L_{\rm bol}$) ratio ($L_{\rm UV}/L_{\rm bol}$) of 0.33 and the scale factor $L_{\rm bol}/(4.5 \times 10^{9} L_{\sun}) = {\rm SFR}/(1 M_{\sun} \ \rm{yr}^{-1})$ that are derived from the starburst population modelling by \citet{Meurer1997}. The derived SFR is then combined with size measurements to calculate the SFI. Note that the SFI is conserved by lensing, since the luminosity increases by the same factor as the area. The SFI is consistent with those of low-redshift analogs such as green peas, and is also seen to be hitting the maximum limit seen in local and high-redshift sources as shown by \citet{Meurer1997,Hathi2008}.

\section{Discussion} 
\label{sec:discuss} 
The difficulties in reproducing the observed line ratios of \JWSTobject\ using photoionization models deserves further attention.   One reasonably likely explanation is that an additional heating source may be present, driving $T_e$ to higher values than photoionization by a starburst tends to achieve.   Either photo-heating by an active galactic nucleus (AGN), or shock heating of the gas, could contribute.  Either of these may have spectroscopic signatures detectable with further {\it JWST} observations.

A second possible class of explanation is density.  The critical density for collisional de-excitation of the \oIIIc\ line is much higher than for \oIII, and for densities $n_e \ga 10^{5} \cm^{-3}$, the observed line ratio can be achieved at considerably lower temperatures \citep[fig.~\ref{fig:osterbrock}; and][]{Netzer1990,Osterbrock2006}.  
While the observed ratio f(\oIIb)/f(\oIIa) does not suggest such high densities, it is difficult to measure confidently given the resolving power of these spectra and the faintness of the \oII\ lines; and moreover it is possible that \oIII\ emission takes place in denser regions of the galaxy than \oII\ emission.  Deeper \oII\ spectra at NIRSpec's highest resolution, or [\ion{S}{2}]6716,6731 measurements using {\it JWST} MIRI spectroscopy, could shed further light on this.

An independent analysis of the line emission in these same galaxies and comparison to nearby analogs was recently published by \citet{Schaerer2022}.  While both their work and ours conclude that the {\it JWST} sources are of low metal abundance and are broadly similar to nearby analogs, there are substantial differences in some of our methodologies.  In particular, they have applied a multiplicative correction (a power law of wavelength) to the spectra to bring the observed Balmer and oxygen line ratios into better agreement with theory, while we have analyzed those line ratios and concluded that within the uncertainties, no correction is required.  As a result, we find a more extreme ratio of \oIIIc\ to \oIII, and consequently a somewhat lower metal abundance in \JWSTobject.

We now return to the larger question of how closely epoch-of-reionization galaxies resemble their best local analogs.  We have shown clear resemblances in their emission-line-dominated rest-frame optical spectra (fig.~\ref{fig:gp_jwst_compare}) and in their small sizes  and high surface brightnesses (fig.~\ref{fig:size}).  The strong emission lines in both sets of galaxies suggest that their luminosities are strongly dominated by young stellar populations, and indeed the hydrogen line equivalent widths in Green Peas require substantial star formation within the last $\sim 5\Myr$.   

On the other hand, low-redshift galaxies almost invariably show underlying, older stellar populations (age $> 1 \Gyr$) when observed in sufficient detail to detect such populations in the presence of younger, brighter stars.  Those underlying populations cannot be present in a universe that is only $0.7\Gyr$ old.  Similarly, detailed abundance ratios in cosmic dawn galaxies may differ from those in older objects, given that entire classes of star may not have had the time to return products of their nuclear burning to the interstellar medium.  Such abundance ratio differences may have an impact on the composition of interstellar dust in these early galaxies.
It may be possible to probe gas-phase abundance ratios in some detail with future JWST NIRSpec and MIRI spectroscopy.  Stellar continuum absorption features may be within reach in a few more years, using 30-meter class telescopes.

We anticipate another important similarity between these galaxies and their local analogs:  The likely presence of strong \lya\ emission.   Among Green Peas, strong \lya\ is nearly ubiquitous \citep{Henry2015,Yang2016,Yang2017}. 
While the present {\it JWST} spectra do not cover \lya\ for these sources, earlier surveys have found \lya\ emitting galaxies up to these same redshifts, based on Lyman break selection \citep{Oesch2015, Zitrin2015}, narrowband imaging \citep[the DAWN survey;][]{Tilvi2020}, or direct slitless spectroscopy \citep[the FIGS survey,][]{Tilvi2016,Larson2018}. And identification of probable \lya\ emitters has been demonstrated not only at low-z \citep{Henry2015,Yang2016} but also in the epoch of reionization using Spitzer photometry to identify the strongest \oIII\ emitters \citep{Roberts-Borsani2016}.   The statistics of \lya\ emission among these early galaxies will be a valuable probe of reionization history \citep{Malhotra2004}.   In particular, if \lya\ is detected in these objects, it will be possible to measure the \lya\ escape fraction by comparison with the Balmer H$\beta$ line, and to estimate what part of \lya\ attenuation is due to the intergalactic medium and what part intrinsic to the galaxy using the Yang relation between velocity offset, dust reddening, and \lya\ escape \citep{Yang2017}.

Finally, these {\it JWST} galaxies are likely giving us our first detailed look at the sources driving cosmological reionization.  Green Peas include a high fraction of galaxies with substantial Lyman continuum escape \citep{Izotov2016,Izotov2018}.  The observable properties of these {\it JWST} targets closely resemble those of Green Peas. Beyond the spectroscopic similarity (fig.~\ref{fig:gp_jwst_compare}), their star formation intensities rank among the highest seen in the Green Pea sample, in a region of paramterer space inhabited by the strongest known Lyman continuum leakers among the Green Peas (figure~\ref{fig:size}).

\section{Conclusions} \label{sec:conclusions}
We have analyzed the rest-frame optical spectra of three epoch-of-reionization galaxies from the {\it JWST} Early Release Observations on the \SMACS\ field.   These objects are all strong line emitters, with spectra reminiscent of nearby Green Pea galaxies, and also of narrowband-selected extreme emission line galaxies at intermediate redshifts.   This result supports earlier conclusions that Green Peas are among the best nearby analogs to high redshift galaxies. 

The highest redshift and most extreme among these galaxies, \JWSTobject\ at $z= 8.495$, has a very low gas phase metallicity, $\zo \approx 6.9 \pm 0.15$ from the $T_e$ method, and $\zo \approx  7.1$ from the \citet{Jiang2019} calibration of the strong line R$_23$ method.  
Its \oIIIc/\oIII\ ratio may demand the presence of a heating mechanism beyond photoheating by a young starburst.
The other two galaxies, \JWSTobjtwo\ at $z=7.664$ and \JWSTobjthree\ at $z=7.659$ have appreciably higher metallicities, in the range $7.7 < \zo < 8.2$.  They demonstrate empirically that galaxy formation and stellar nucleosynthesis can achieve a metal abundance comparable to the Magellanic Clouds \citet{Russell1992} within just 700 Myr after the Big Bang.

All three of these galaxies share the compact sizes and high surface brightnesses that characterize \lya\ emitting galaxies across a wide range of redshifts, from $z\sim 6.5$ down to $z\sim 0$ \citep{Malhotra2012,Kim2020,Kim2021}.

Low-redshift analogs for epoch-of-reionization galaxies have been of tremendous value in recent years, because they allowed us to study in nearby objects properties that could not be directly studied in faint, redshifted galaxies at Cosmic Dawn.  As we have demonstrated, {\it JWST} now enables direct measurements of the many physically interesting quantities that can be derived from rest-frame optical emission lines.  These will ultimately include metallicity, temperature, ionization paramter, density, and gas pressure.
Despite this, the importance of local analogs remains.  Some properties remain beyond reach at high redshift due to sensitivity.   For exmaple, radio emission from atomic gas in single galaxies can be studied in green peas \citep{Kanekar2021,Purkayastha2022} but is far beyond reach in the early universe.  Other important measurements are more fundamentally precluded at high redshift, notably including the escape fraction of ionizing radiation,  which cannot be effectively observed at $z \ga 4$ due to absorption by residual neutral gas in the intergalactic medium.  The detection of Lyman continuum escape fractions of tens of percent, and even $>50\%$ in some Green Pea galaxies,
is a key ingredient in understanding reionization sources \citep{Izotov2018,Flury2022,Flury2022b}. Thus, {\it JWST} now equips us to establish the validity of local analog populations with unprecedented detail and confidence, opening the way for further progress using both the most distant and the closest young galaxies.

\begin{acknowledgments}
We thank the {\it JWST} team--- {\it all} of you--- for making this possible.  

This work is based in part on observations made with the NASA/ESA/CSA James Webb Space Telescope. The data were obtained from the Mikulski Archive for Space Telescopes at the Space Telescope Science Institute, which is operated by the Association of Universities for Research in Astronomy, Inc., under NASA contract NAS 5-03127 for JWST. These observations are associated with program \#2736.
The authors acknowledge the JWST ERO team for developing their observing program with a zero-exclusive-access period.

This work has made use of public data from the Sloan Digital Sky Survey (SDSS).
Funding for the SDSS and SDSS-II has been provided by the Alfred P. Sloan Foundation, the Participating Institutions, the National Science Foundation, the U.S. Department of Energy, the National Aeronautics and Space Administration, the Japanese Monbukagakusho, the Max Planck Society, and the Higher Education Funding Council for England. The SDSS Web Site is http://www.sdss.org/.

The SDSS is managed by the Astrophysical Research Consortium for the Participating Institutions. The Participating Institutions are the American Museum of Natural History, Astrophysical Institute Potsdam, University of Basel, University of Cambridge, Case Western Reserve University, University of Chicago, Drexel University, Fermilab, the Institute for Advanced Study, the Japan Participation Group, Johns Hopkins University, the Joint Institute for Nuclear Astrophysics, the Kavli Institute for Particle Astrophysics and Cosmology, the Korean Scientist Group, the Chinese Academy of Sciences (LAMOST), Los Alamos National Laboratory, the Max-Planck-Institute for Astronomy (MPIA), the Max-Planck-Institute for Astrophysics (MPA), New Mexico State University, Ohio State University, University of Pittsburgh, University of Portsmouth, Princeton University, the United States Naval Observatory, and the University of Washington.
\end{acknowledgments}

\bibliography{refs}{}
\bibliographystyle{aasjournal}



\end{document}